# Dreidel Fairness Study

Robert Nemiroff
&
Eva Nemiroff

Email: nemiroff@mtu.edu

## Abstract

Are dreidels fair? In other words, does the average dreidel have an equal chance of turning up any one of its four sides? To explore this hypothesis, three different dreidels were each spun hundreds of times with the number of occurrences of each side recorded. It was found that all three dreidels tested -- a cheap plastic dreidel, an old wooden dreidel, and a dreidel that came embossed with a picture of Santa Claus -- were not fair. Statistically, for each dreidel, some sides came up significantly more often than others. Although an unfair dreidel does not necessarily make the game itself unfair, it is conjectured that hundreds of pounds of chocolate have been distributed during Chanukah under false pretenses.

## Introduction

A game popular with many children during Chanukah is called Dreidel. The game involves the spinning of a four-sided top called, not-coincidentally, a dreidel. The word "dreidel", as used in English, derives from the word dreyen which translates "to spin" in Yiddish [1].

The rules of Dreidel may vary, but invariably involve people spinning a dreidel and examining which side turns up [2]. A common unstated assumption is that each side has an equal chance of ending face-up. Depending on the side showing, a player may be required to place coins in or out of a central pot. In the modern experience of the authors, these coins are typically made of foil-wrapped chocolate and referred to as "Chanukah gelt".

Over the years, the lead author, surely among many others, has been known to wonder aloud why certain sides of a dreidel seem to turn up more often than others. Nevertheless, to the best of our knowledge, no formal study of the fairness of dreidels has been done. Therefore, the researchers have performed one such study and report the results here.

Three dreidels were procured for this study. The first was a small, common, plastic dreidel that was chosen randomly from 25 similar dreidels ordered online from Amazon.com in 2015 January. Given that all 25 dreidels together cost $8.48, this is called here the "cheap plastic dreidel". The second dreidel was small and wooden and just found around the researcher's house. Although this dreidel appeared familiar to everyone asked, no one could recall its origin. This dreidel is here dubbed the "old wooden dreidel". The third dreidel was discovered serendipitously online while ordering the first dreidel. It was relatively large and in the place of more-typical Hebrew letters, it had pictures of Santa Claus, a Christmas Tree, a snowman, and a candy cane. It was not known to the researchers previously that dreidels with such decorations existed, and, despite exceeding a predetermined price point, was so intriguing that that it was ordered and soon referred to as the "Santa dreidel". All three dreidels are shown in Figure 1.

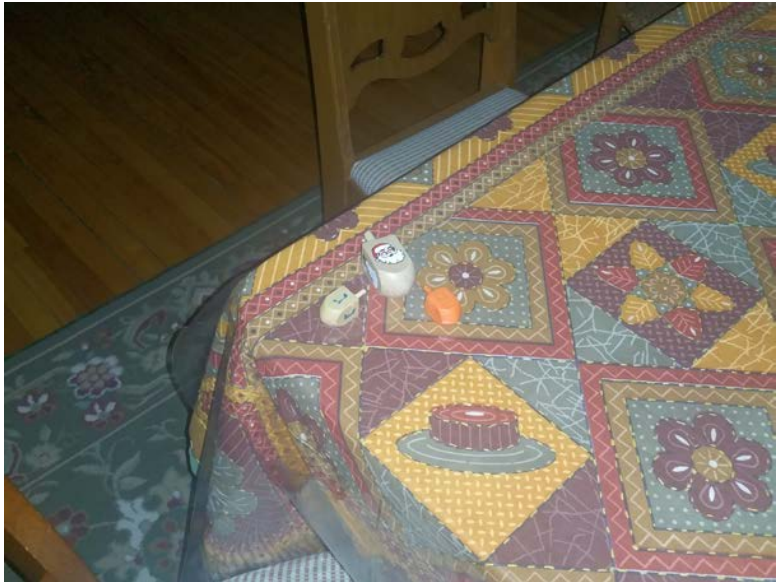
Figure 1: The three dreidels used this study.

**Procedure**

Both researchers, not-coincidentally father and daughter, spun all three dreidels, spun them at different times over 11 days in February and March 2015, and spun them on two different surfaces. Complaints of boredom were disfavored. The surface of the table in the TV room was relatively hard, while the surface of dining room table was comparatively soft as it was covered with a thick plastic tablecloth. Each spin was a "real spin" in that the dreidel revolved at least several times and it was not initially obvious to the spinner which side would finish face-up. The researchers were not trying to create any result -- it was really unknown to them whether each dreidel was fair, and they were curious to find out. Consecutive spins were conducted in a simple manner similar to spins made and observed at numerous previous Chanukah celebrations. After each spin, a researcher would record the result with a tick mark on a piece of paper. These tick marks were later counted up and transcribed on the family computer. Spins that caused the dreidel to fall off the table were respun on the table. This occurrence, although potentially a point of arduous debate, was deemed rare enough so as to not significantly bias the results.

## Results

None of the dreidels tested were fair. After 2,550 spins, a straightforward statistical analysis involving the Chi-Squared parameter showed a $6.7 \times 10^{-6}$ chance that the spins produced by the cheap plastic dreidel were consistent with a fair dreidel. And that dreidel was the most fair of the three! The old wooden dreidel was the least fair, with a miniscule $4.8 \times 10^{-48}$ chance of being unbiased, while the Santa dreidel's spins were in the middle with an also tiny $4.6 \times 10^{-28}$ chance of being equitable. Basic results are summarized in Table 1.

Table 1

| Dreidel | ג or Santa | נ or candy | ש or פ or tree | ה or snow-man | Spins | $\chi^2$ | $P(\geq \chi^2)$ |
|---|---|---|---|---|---|---|---|
| Santa | 109 | 302 | 134 | 255 | 800 | 130 | $4.6 \times 10^{-28}$ |
| Cheap plastic | 311 | 243 | 196 | 250 | 1000 | 26.7 | $6.7 \times 10^{-06}$ |
| Old wooden | 52 | 275 | 126 | 297 | 750 | 223 | $4.8 \times 10^{-48}$ |

A more detailed statistical analyses showed that results did not significantly depend on who did the spinning and that the surface the dreidel was spun on had little effect on the cheap plastic and Santa dreidels. Oddly, the nature of the surface did appear to be a significant factor for the old wooden dreidel. Although the same sides of this dreidel were preferred on both surfaces, the chance that the two surfaces had the same effect on this dreidel was less than $3.39 \times 10^{-4}$.

## Conclusions

The results reported are formal statistical confidence levels. It is quite possible that significant systematic errors were operating, such as non-random spinning techniques, or that one side of a dreidel was unusually sticky due to someone -- a person not here named -- eating candy, getting sticky fingers, and then touching dreidels but not reporting it. Although such systematic effects could well reduce the extremely high statistical confidence levels reported, it is expected that such effects were not determinative. Therefore, it is suggested that the reported results do clearly indicate that the dreidels tested were not fair, and by implication, that many common dreidels in operation at Chanukah celebrations were similarly unfair.

Now just because the dreidel itself was unfair does not mean necessarily that the game itself was unfair. Assuming that all game players have an equal number of chances to spin the dreidel, the bias of the dreidel should affect all players equally. However, were players to have an unequal number of spins [3,4], or if a player insisted on using their own dreidel, a bias in the dreidel could affect the fairness of the game. At minimum, an unfair spinner creates a false pretense to those believing that the spinner is fair.

Taking this result as indicating that most dreidels are inherently unfair, how much chocolate has been distributed, during Chanukah celebrations, under this false pretense? To estimate this several assumptions have been made. These include that all players have assumed, to date, that all dreidels were fair; that at least half of all dreidel spins were unfair; that at least one ounce of chocolate coins (gelt) was typically used for Dreidel at each at each Hanukkah celebration; that there have been at least 10,000 such celebrations per year; and finally that chocolate has been significantly involved these celebrations for at least 20 years. Given these assumptions, then over 100,000 ounces -- over 1,000 pounds -- of chocolate has been distributed in deceiving circumstances.

What should be done? Although it is arguably outside the purview of this scientific study to untangle implied moral dilemmas, while we have your attention we will make a few suggestions. First, forgiveness: in the spirit of giving, we suggest amnesty for any past Dreidel winners so that any chocolate winnings, for example, are not asked to be returned. Next, overlooking the moral uncertainty inherent in teaching children the supposed joys of gambling, we suggest that Dreidel games be continued as a positive family and community building activity. Since the children are going to end up eating most the chocolate anyway, distributing some of it during a game might be considered more fun than just giving it to them one Chanukah night and saying "here". Last, even after the chocolate is distributed, it should be pointed out that part of the importance of Chanukah is not in who wins at a childish game, but rather that people realize that the reason that you lost, this year, was that the dreidel being used was unfairly lopsided.

______________________________